\documentclass[sigconf, nonacm]{acmart}

\usepackage{algorithmic}
\usepackage{graphicx}
\usepackage{textcomp}
\usepackage{xcolor}
\usepackage{url}

\usepackage{siunitx}
\DeclareSIUnit{\pixel}{px}

\usepackage[nolist]{acronym}

\usepackage{float}
\usepackage{placeins}

\usepackage{subcaption}

\usepackage{tikz}
\usetikzlibrary{calc}
\usetikzlibrary{positioning}

\usepackage{balance}

\usepackage{ifthen}
\newboolean{usetikzexternalize}
\setboolean{usetikzexternalize}{true}

\newlength{\halflinewidth}
\newcommand{\mysubsubsubsection}[1]{
\vspace{1.5pt plus .5pt minus .5pt}
\noindent\textit{\textbf{#1\enskip}}
}

\newcommand{\dwelltime}[0]{\Delta t_{px}}

\newboolean{cameraready}
\setboolean{cameraready}{true}

\newboolean{arxiv}
\setboolean{arxiv}{true}

\ifthenelse{\boolean{arxiv}}{
	\settopmatter{printacmref=false, printccs=false, printfolios=true} %
	\hypersetup{colorlinks,
		linkcolor=ACMPurple,
		citecolor=ACMPurple,
		urlcolor=ACMDarkBlue,
		filecolor=ACMDarkBlue}
}{
	\settopmatter{printacmref=true}
}

\ifthenelse{\boolean{usetikzexternalize}}{
	\usetikzlibrary{external}
	\tikzexternalize[prefix=figs/ext/]
	
}{}

\ifthenelse{\boolean{cameraready}}{
	\ifthenelse{\boolean{arxiv}}{
	}{
	}
	
}{
}

\begin{acronym}
	\acro{ht}[HT]{hardware Trojan}
	\acro{fpga}[FPGA]{field-programmable gate array}
	\acro{asic}[ASIC]{application-specific integrated circuit}
	\acro{ip}[IP]{intellectual property}
	\acro{ic}[IC]{integrated circuit}
	\acro{fb}[FB]{feedback}
	\acro{bga}[BGA]{ball grid array}
	\acro{clb}[CLB]{configurable logic block}
	\acro{lut}[LUT]{lookup table}
	\acro{ff}[FF]{flip-flop}
	\acro{le}[LE]{logic element}
	\acro{lc}[LC]{logic cluster}
	\acro{il}[IL]{interface logic}
	\acro{mux}[MUX]{multiplexer}
	\acroplural{mux}[MUXes]{multiplexers}
	\acro{soc}[SoC]{system on a chip}
	\acroplural{soc}[SoCs]{systems on a chip}
	\acro{llsi}[LLSI]{laser logic state imaging}
	\acro{eofm}[EOFM]{electro-optical frequency mapping}
	\acro{lvi}[LVI]{laser voltage imaging}
	\acro{eop}[EOP]{electro-optical probing}
	\acro{lvp}[LVP]{laser voltage probing}
	\acro{fa}[FA]{failure analysis}
	\acro{pcb}[PCB]{printed circuit board}
	\acro{sil}[SIL]{solid immersion lens}
	\acro{sca}[SCA]{side-channel analysis}
	\acro{hil}[HIL]{high-power incoherent light source}
	\acro{na}[NA]{numerical aperture}
	\acro{dut}[DUT]{device under test}
	\acroplural{dut}[DUTs]{devices under test}
	\acro{ide}[IDE]{integrated development environment}
	\acro{nvm}[NVM]{non-volatile memory}
	\acro{em}[EM]{electromagnetic}
	\acro{pe}[PE]{photon emission}
	\acro{nir}[NIR]{near-infrared}
	\acro{fib}[FIB]{focused ion beam}
	\acro{sem}[SEM]{scanning electron microscopy}
\end{acronym}

\begin{document}
\ifthenelse{\boolean{cameraready}}{
	\ifthenelse{\boolean{arxiv}}{
	}{
		\fancyhead{}
	}
}

\title[Trojan Awakener: Detecting Dormant Malicious Hardware Using Laser Logic State Imaging (Extended Version)]{Trojan Awakener: Detecting Dormant Malicious Hardware Using Laser Logic State Imaging (\underline{Extended Version})}
\titlenote{For remarks on the extended version, see the last paragraph of Section~\ref{sec:introduction}}

\author{Thilo Krachenfels}
\affiliation{
	\institution{Technische Universität Berlin - SECT}%
	\country{Berlin, Germany}
}
\email{tkrachenfels@sect.tu-berlin.de}

\author{Jean-Pierre Seifert}
\affiliation{
	\institution{Technische Universität Berlin - SECT}%
	\country{Berlin, Germany}
}
\email{jpseifert@sect.tu-berlin.de}

\author{Shahin Tajik}
\affiliation{
	\institution{Worcester Polytechnic Institute}%
	\country{Worcester, USA}
}
\email{stajik@wpi.edu}

\renewcommand{\shortauthors}{Krachenfels, Seifert, and Tajik}

\begin{abstract}	
	The threat of \acp{ht} and their detection is a widely studied field.
	While the effort for inserting a Trojan into an \ac{asic} can be considered relatively high, especially when trusting the chip manufacturer, programmable hardware is vulnerable to Trojan insertion even after the product has been shipped or during usage.
	At the same time, detecting dormant \acp{ht} with small or zero-overhead triggers and payloads on these platforms is still a challenging task, as the Trojan might not get activated during the chip verification using logical testing or physical measurements.
	In this work, we present a novel Trojan detection approach based on a technique known from \ac{ic} failure analysis, capable of detecting virtually all classes of dormant Trojans.
	Using \ac{llsi}, we show how supply voltage modulations can awaken inactive Trojans, making them detectable using laser voltage imaging techniques.
	Therefore, our technique does not require triggering the Trojan.
	To support our claims, we present three case studies on \SI{28}{\nm} and \SI{20}{\nm} SRAM- and flash-based \acp{fpga}.
	We demonstrate how to detect with high confidence small changes in sequential and combinatorial logic as well as in the routing configuration of \acp{fpga} in a non-invasive manner.
	Finally, we discuss the practical applicability of our approach on dormant analog Trojans in ASICs.
\end{abstract}
\acresetall


\maketitle

\section{Introduction \label{sec:introduction}}

Due to their reconfigurability, high performance, and a short time to market, programmable hardware, especially \acp{fpga}, have become the default solution in many fields.
One of the main strengths of \acp{fpga} compared with \acp{asic} is that the hardware configuration can be updated and even reprogrammed during runtime.
At the same time, the demand for security increases as more and more security-critical systems are based on electronics.
Therefore, malicious modifications of the design, referred to as \acp{ht}, endanger the security of many applications.
On \acp{fpga}, a Trojan might be inserted after manufacturing and testing, i.e., in the untrusted field~\cite{ng2015integrated,roy2018conflicted}, for instance, by altering the entire configuration (known as bitstream) or via partial reconfiguration.
Particularly if the chip foundry can be trusted, this depicts a much more powerful threat model than for ASICs.
Most security-critical FPGAs rely on bitstream encryption and authentication to avoid such Trojan insertions.
However, these protection schemes have shown to be vulnerable to various physical~\cite{moradi_improved_2016,tajik2017power,lohrke_key_2018,hettwer2021side} and mathematical attacks~\cite{ender2020unpatchable}, leaving them susceptible to tampering.
Consequently, in critical applications, where the chip is deployed in an untrusted field or could be accessed by untrusted parties, it should be possible to check the integrity of the hardware.

Integrity checking of running applications on \acp{fpga} in the field faces mainly two obstacles.
First, while checking the configuration against a golden bitstream would reveal tampering (as proposed in~\cite{zhang_thwarting_2019}), it is not possible in many cases.
In several defense/aerospace applications, where flash-based FPGAs~\cite{microchip_polarfire_security} or SRAM-based FPGAs with preemptive decryption key zeroization~\cite{xilinx_tamper_resistance} are deployed, no bitstream (encrypted or unencrypted) is available to the hardware testing engineer in the field for verification.
In these cases, the configuration is stored inside the chip and bitstream readback is not possible.
Even if the bitstream is available, analyzing the unencrypted bitstream is not an option since the circuit and the secret keys for bitstream decryption should be unknown even to the testing engineer.
Moreover, the same bitstream can be encrypted with various keys for different FPGAs, and therefore, comparing encrypted bitstreams to each other for tampering detection might also not be feasible.

Second, while early \acp{ht} had logic triggers that could be activated by logical testing~\cite{salmani2009new} under some circumstances, recently proposed \acp{ht} are classified as \emph{stealthy} or \emph{dormant}.
In other words, the Trojan payload reacts only under extremely rare conditions, for instance, in a particular temperature, supply voltage, or frequency range~\cite{ender_first_2017} or after a certain amount of specific events have occurred~\cite{yang_a2_2016}.
Furthermore, under operational and testing conditions, a dormant Trojan tries to hide from physical inspection or side-channel analysis, e.g., by leveraging analog components~\cite{yang_a2_2016}, manipulating only the dopant level of the chip~\cite{becker_stealthy_2013}, or changing only the routing configuration on programmable hardware~\cite{ender_first_2017}.

Several approaches based on \ac{sca} for detecting such dormant \acp{ht} have been proposed in the literature~\cite{song_marvel_2011, stellari_verification_2014, duncan_flats_2019, nguyen_creating_2019, adibelli_field_2020, he_golden_2020, stern2020sparta, zhou_hardware_2020}.
However, they all face severe limitations regarding resolution and the capability to detect all types of \acp{ht}.
For instance, approaches using \ac{em} backscattering side-channels are naturally limited by their resolution and can only detect larger malicious design changes~\cite{nguyen_creating_2019, adibelli_field_2020}.
Furthermore, these approaches can reliably detect dormant Trojans only with a high rate of false positives.
One technique that provides higher resolution is optical probing, where the chip is scanned through its backside with a laser, and the reflected light is analyzed.
However, the reported approach based on \ac{eofm}~\cite{stern2020sparta} is limited to detecting malicious modifications only in the sequential logic, and thus, Trojans that solely consist of combinatorial logic stay undetected.

A new optical probing technique that has recently been leveraged in the hardware security field is called \emph{\ac{llsi}}~\cite{krachenfel_realworld_2021}.
It is an optical probing technique that can extract the logic states of single transistors, and therefore, more complex logic gates or memory cells~\cite{niu_laser_2014}.
In \ac{llsi}, the chip's supply voltage is modulated, which causes the light reflection originating from a laser scanning irradiation to be modulated as well.
The modulation amplitude is dependent on the carrier concentration present in the silicon, for instance, inside the channel of a transistor.
Consequently, the \ac{llsi} signal is highly data-dependent and provides a practically unlimited number of electro-optical probes.
Hence, it should be possible to extract the configuration of an FPGA's logic fabric using \ac{llsi}, especially because the configuration is held in memory cells distributed over the chip.
The logic state of these cells controls the functioning of \acp{lut}, \acp{mux}, and pass transistors in switch boxes.
In this work, we try to clarify \emph{if small dormant \acp{ht} on state-of-the-art FPGAs -- consisting of combinatorial or sequential logic -- can be detected by applying \ac{llsi}.}

\vspace{2mm}
\noindent \textbf{Our contribution.}
We indeed positively answer the above question.
First, we present how \acs{llsi} allows us to capture the state of every transistor of the logic fabrics of SRAM- and flash-based FPGAs.
Based on this, we demonstrate how to partially reverse-engineer the FPGA's configuration, including the detection of changes in a single \ac{lut}.
Second, we show how this new approach can detect small and dormant \acp{ht} on FPGAs.
Stimulating all transistors with the power supply modulation awakens maliciously modified hardware, from which we then can take a snapshot.
Therefore, the Trojan can be inactive/dormant, as our approach does not rely on any switching activity on the chip.
For detecting \acp{ht}, we first capture a reference snapshot of the FPGA's logic fabric in the trusted field -- when the design is known to be Trojan-free.
Later, to check if the design has been altered, we capture a snapshot of the logic fabric and compare it to the reference.
We show that the high resolution of optical probing allows detecting small changes of the configuration, down to changes in a single combinatorial gate.

Our approach can be applied \emph{non-invasively} since almost all current \acp{fpga} are available in flip-chip packages allowing easy access to the silicon backside.
To validate our claims, we present three case studies on SRAM- and flash-based \acp{fpga} from Xilinx (\SI{28}{\nm} and \SI{20}{\nm} technology) and Microchip (\SI{28}{\nm} technology), respectively.
Although our experiments are focused on \acp{fpga}, we discuss why \ac{llsi} is applicable for analog \ac{ht} detection on \acp{asic}.

\vspace{2mm}
\noindent \textbf{Remarks on the extended version.}
The original version of this work has been presented at the \textit{Attacks and Solutions in Hardware Security (ASHES 2021)} workshop~\cite{krachenfel_trojan_2021}.
The version at hand contains the following additional and revised content: i) the investigation of a new target device manufactured in a \SI{20}{\nano\m} technology, including setup, results, and discussions; ii) a more thorough explanation and discussion of the experimental setup, especially regarding the \ac{llsi} modulation frequencies; iii) a detailed discussion of how to prepare a real-world device that should be investigated using the presented \ac{ht} detection approach; and iv) additional figures depicting the experimental setup.

\section{Background \label{sec:background}}

\subsection{Hardware Trojans}

\subsubsection{Properties and Taxonomy}
The term \acf{ht} includes a wide range of malicious circuit modifications which, for instance, try to leak sensitive information through side-channels, implement kill-switches and backdoors, or enforce faulty computations.
\Acp{ht} can be characterized by their physical properties (e.g., type and size of modifications), activation characteristics (i.e., trigger source and frequency), and action characteristics (i.e., which goal the \ac{ht} serves)~\cite{wang_detecting_2008}.
As diverse as the different types of \acp{ht} are, so are the potential entities that might introduce the malicious modifications~\cite{bhunia_hardware_2014}.
During the development and production of \acp{ic}, weak points include third-party \ac{ip} cores, malicious design tools, and mask layout or doping concentration modifications~\cite{becker_stealthy_2013a} by untrusted foundries.
The platform TrustHub~\cite{shakya_benchmarking_2017} provides several design-level \ac{ht} benchmarks, primarily available as gate-level descriptions.
TrustHub provides access to the automatically generated \ac{ht} benchmarks presented in~\cite{cruz_automated_2018} that alter existing circuit designs by inserting malicious logic gates.

Programmable hardware devices, like \acp{fpga}, are less prone to \emph{production-based} \ac{ht} insertion than \acp{asic}.
On the other hand, due to their reconfigurability, they provide the possibility for malicious modifications even after the product has been shipped to the user.
It has been shown that the key used for encrypting the bitstream on recent SRAM-based \acp{fpga} can be extracted using \ac{sca} techniques~\cite{moradi_improved_2016,tajik2017power,lohrke_key_2018,hettwer2021side}.
With the extracted key at hand, the bitstream can be decrypted, modified, and stored as a replacement for the original bitstream~\cite{duncan_flats_2019}.
Although bitstream extraction from flash-based \acp{fpga} might not be possible, the adversary could still be able to reprogram certain parts of the configuration or even replace the entire chip containing her malicious version of the design.

\subsubsection{Hardware Trojans on FPGAs}
While generic Trojans, such as backdoors, can be implemented on both \acp{asic} and programmable hardware, a few \acp{ht} especially tailored to \acp{fpga} have been proposed.
For instance, Jacob et al. have proposed an approach that exploits shared resources between the programmable logic and the embedded microcontroller on an \ac{fpga} \ac{soc}~\cite{jacob_compromising_2017,jacob2017break}.
By hidden functionalities in an \ac{ip} design block, the programmable logic can access and manipulate shared memory locations used for storing sensitive information like cryptographic keys.
Ender et al. have proposed a Trojan that is solely based on minor timing modifications on the chip~\cite{ender_first_2017}.
They show that by operating the chip with modified signal paths at a specific frequency, the data masking scheme protecting against side-channel analysis attacks is not functional anymore, allowing the extraction of the secret key used in the protected algorithm.
They show that on an \ac{fpga}, longer signal paths can be realized by instantiating route-thru \acp{lut}, or by modifying the routing in the switch boxes, which results in zero overhead in resource usage, and therefore, is hard to detect.
In another effort, Roy et al. showed that the reconfigurable LUTs could be exploited to realize \acp{ht} with zero payload overheads~\cite{roy2018conflicted}.
Finally, Ng et al. demonstrated that integrated sensors inside FPGAs could be deployed as Trojan triggers~\cite{ng2015integrated}.

\subsubsection{Detection of Hardware Trojans -- Related Work}\label{sec:fpga_tojan_detection}
As already mentioned in Section~\ref{sec:introduction}, \ac{ht} detection on FPGAs cannot always be carried out by checking or comparing bitstreams.
Therefore, most of the \ac{ht} detection techniques use different kinds of physical measurements and side-channel information obtained from the chip.
Optical chip backside reflectance imaging~\cite{zhou_hardware_2020}, \ac{sem} imaging~\cite{vashistha2018trojan}, or \ac{fib} imaging~\cite{sugawara2014reversing} are not suitable for detecting \acp{ht} on \acp{fpga}, because the physical design and layout of the chip do not depend on the actual programmed functionality.
\Ac{sca} techniques, such as power analysis, \ac{em} analysis~\cite{he_golden_2020}, or backscattering analysis~\cite{nguyen_creating_2019, adibelli_field_2020}, can be used for all types of \acp{ic}.
By applying different clustering algorithms, the Trojan-infected chips can be separated from the non-infected chips, often without the need of a golden chip, i.e., a chip which is known to be Trojan-free.
However, these techniques only offer a limited resolution, which requires the Trojan trigger logic to consist of a minimum number of gates or being separated from its input signals to a certain extent~\cite{nguyen_creating_2019}.
Furthermore, the clustering does only work if the set of samples contains at least one non-infected device. %

\Ac{sca} techniques offering higher resolution include approaches that observe the chip's operation through the silicon backside, which is transparent to \ac{nir} light.
For instance, \ac{pe} analysis can be used to compare dynamic and static emissions with the chip layout~\cite{stellari_verification_2014} or emissions from a golden chip~\cite{song_marvel_2011}.
Furthermore, adding oscillators with inputs from the design that act as beacons can facilitate the detection of tampering attempts, especially when cheaper infrared imaging is used~\cite{duncan_flats_2019}.
However, such an approach increases the resource consumption of the design considerably in many cases and might not be able to detect all possible changes in \ac{lut} configurations.
One approach providing higher resolution and better localization capabilities is optical probing.
The authors of \cite{stern2020sparta} have demonstrated that using an optical probing technique, all \acp{ff} used in the hardware design can be located and mapped to the intended design from the \ac{fpga} \ac{ide}.
In this way, malicious changes in the sequential logic can be detected reliably and in a non-invasive fashion, if the chip is packaged as flip-chip.
However, combinatorial logic can not be detected using that approach, which is the major downside of the approach.

\subsection{Field-Programmable Gate Arrays (FPGAs)}
The heart of an FPGA is its configurable logic fabric, consisting of an array of small configurable logic elements %
containing \acfp{lut} and \acfp{ff} for implementing combinatorial and sequential logic respectively.
Configurable routing resources interconnect these blocks.
Together with on-chip memories and input/output capabilities, such as transceivers, the designer can implement virtually every functionality on the FPGA.
To add the software configurability of processors to FPGAs, vendors offer soft processor cores, and recently even \acp{soc} containing both ASIC processors and an FPGA logic fabric, connected by an effective interconnection network.

\begin{figure}[tb]
	\centering
	\includegraphics[width=\linewidth]{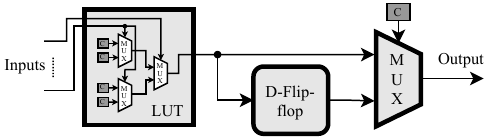}
	\caption{Simplified schematic of an FPGA logic block. \acp{lut} and \acp{mux} are controlled by configuration memory cells.\label{fig:background_le}}
\end{figure}

Although the logic fabric architecture differs between manufacturers, the building blocks are multi-input \acp{lut} for combinatorial logic, \acp{ff} for sequential logic, and \acp{mux} for signal routing, see Fig.~\ref{fig:background_le}.
The two main configuration storage types for FPGAs are volatile SRAM-based and non-volatile flash-based memories.

\subsubsection{SRAM-based}
The dominating manufacturers for \acp{fpga} are Xilinx (acquired by AMD) and Intel (formerly Altera), with a combined share of more than 85\%~\cite{dougblack_xilinx_2019}.
Both of them focus on SRAM-based \acp{fpga}.
The advantage of using SRAM as memory technology is that the chip can be manufactured with cutting-edge chip technologies, which allows for higher logic densities.
Due to the volatile nature of SRAM cells, the \ac{fpga}'s configuration is lost after every power-down.
Therefore, the configuration (the bitstream) must be stored in external memory and loaded upon every reboot by the \ac{fpga}'s configuration fabric.
This fabric decrypts the configuration and loads it into the distributed SRAM cells on the chip, which determine the behavior of \acp{lut}, \acp{mux}, and routing transistors.
One advantage of the volatile configuration storage is the possibility to partially reconfigure the logic fabric during runtime. %

\subsubsection{Flash-based}
Flash-based FPGAs are offered mainly by Microchip (formerly Microsemi) and Lattice Semiconductor, with a combined market share smaller than 12\%~\cite{dougblack_xilinx_2019}.
The main advantage of flash-based FPGAs over SRAM-based FPGAs is their lower power consumption.
Further, the configuration is stored in a non-volatile way in distributed flash cells.
One reason for the lower power consumption is that flash cells consist of fewer transistors than SRAM cells and do not need to be powered for retaining their value.

\subsection{Laser-Based Logic Readout\label{sec:background:llsi}}

\begin{figure}[tb]
	\centering
	\includegraphics[width=\linewidth]{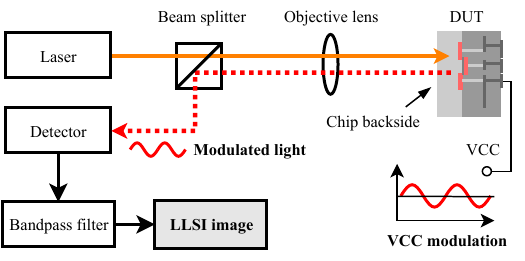}
	\caption{Schematic of LLSI image acquisition. The DUT is scanned with a laser through the chip backside; due to a power supply (VCC) modulation, the reflected light is modulated, which can be detected.\label{fig:background_llsi}}
\end{figure}

\subsubsection{Technique}
Optical probing is a powerful approach known from \ac{ic} \ac{fa}.
A laser is pointed on the chip's backside, and switching activity causes the reflected laser light to modulate.
More specifically, mainly the concentration of free carriers distinguishes the refraction and absorption of the laser light in silicon.
When the laser scans the device and the reflected signal is fed through a bandpass filter set to a frequency of interest, all areas on the chip switching at a frequency of interest can be detected. 
The corresponding technique is called \acf{eofm} or \acf{lvi}.

Using classical \ac{eofm}, only periodically switching elements on the chip can be detected.
The static logic state of circuits, however, can be captured using \acf{llsi}, which was introduced as an extension to \ac{eofm}~\cite{niu_laser_2014}.
The main idea behind \ac{llsi} is to stop the clock and induce a periodic frequency into the entire logic by modulating the power supply, see Fig.~\ref{fig:background_llsi}.
This causes the free carrier concentrations to vary periodically, e.g., in the channel of transistors or in capacitors.
This, in turn, modulates the reflected light, which can be detected using \ac{eofm}.
Transistors that are switched on (low-ohmic channel) can thus be distinguished from transistors that are switched off (high-ohmic channel).

\subsubsection{Related Work}
\Ac{llsi} has been used in the hardware security field to extract the values stored in SRAM cells or \acp{ff}.
The authors of \cite{krachenfel_realworld_2021} demonstrated that the \ac{ff} content of an \ac{fpga} manufactured in a 60\,\si{\nano\meter} technology can be extracted using \ac{llsi}.
Using classical image recognition techniques, they show that the content can be extracted in an automated fashion.
In \cite{krachenfel_automatic_2021}, the authors demonstrate that a key stored in the SRAM of a microcontroller can be extracted using \ac{llsi} combined with deep learning techniques without the need to reverse-engineer the chip's layout.
To the best of our knowledge, \ac{llsi} has neither been used to extract an \ac{fpga}'s logic fabric configuration nor to detect \acp{ht}.

\section{Approach \label{sec:approach}}
\begin{figure}[tb]
	\centering
	\includegraphics[width=\linewidth]{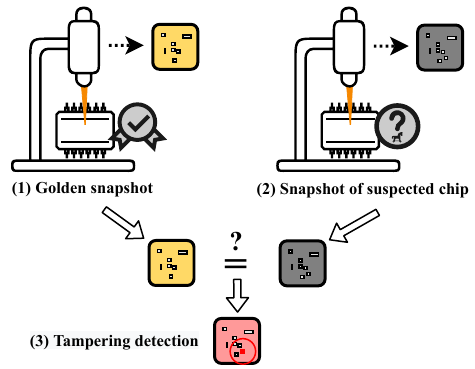}
	\caption{Approach for detecting tampering with the FPGA logic fabric configuration.
		\label{fig:approach}}
\end{figure}
In our scenario, the supply chain from the finished product to the field can not be trusted.
In other words, an adversary might replace or change the device’s functionality after it has left the trusted design house. In such a scenario, the highest efforts are paid to detect malicious hardware, e.g., in military, space, and aircraft applications.
Although \ac{llsi} can capture the states of transistors and memory cells in all \acp{ic} %
, our goal in this work is to apply \ac{llsi} for creating snapshots of the logic fabric in \acp{fpga}.
To do so, we need to modulate the supply voltage of the logic under test, in our case, of the logic fabric, see Section~\ref{sec:background:llsi}.
Furthermore, we need to halt the clock of the \ac{fpga}.
To test if the \ac{fpga}'s configuration manifests in the hardware snapshots, we configure the logic fabric in different ways, for instance, by altering the configuration of \acp{lut} and the routing.
We then compare the snapshot images to see if the changed configuration can be detected and at which location the change has occurred.

Once different configuration changes can be detected, the knowledge can be used to also detect malicious modifications on the chip, see Fig.~\ref{fig:approach}.
In our approach, we create a snapshot of the original Trojan-free design, also known as golden design, in the trusted design house (1).
It typically will be necessary to create multiple snapshots to cover the entire logic fabric area with high resolution.
We then assume a malicious entity 
that inserts a Trojan into the \ac{fpga} configuration of the product.
Before using the final product in a security-critical application, the integrity of the \ac{ic} should be certified.
For this, we create a snapshot of the suspected chip (2).
To eliminate the chance of any tampering, we compare the golden snapshot with the current snapshot (3).
For comparing the snapshots, subtracting the images %
might be helpful.
If there are differences, this indicates that the configuration has been altered, and the chip is not trustworthy.
It should be noted that the state of the \acs{fpga} in step (1) and (2) should be the same, i.e., the clock should be stopped in the same cycle.
We expect our approach to work on both SRAM- and flash-based FPGAs.

\mysubsubsubsection{SRAM-based FPGAs}
SRAM-based FPGA configuration takes place by configuring LUTs and global/local routing via SRAM cells.
In the end, all configuration SRAM cells do control \acp{mux}, which consist of pass transistors.
Since \ac{llsi} can extract the logic states of CMOS transistors, the \ac{fpga}’s entire configuration should be extractable -- given a sufficiently high optical resolution.

\mysubsubsubsection{Flash-based FPGAs}
The configuration of flash-based \acp{fpga} is stored in dedicated flash cells, which are distributed over the chip.
They control the \acp{lut} and global/local routing using multiplexers, which, like in SRAM-based \acp{fpga}, consist of pass transistors.
Therefore, also the configuration of flash-based \acp{fpga} should be extractable using \ac{llsi}.
If the flash cells are supplied by another voltage rail, it might be possible to see a configuration dependency by modulating that rail.

\section{Experimental Setup \label{sec:setup}}
This section first presents our measurement setup, followed by the \acp{dut} and their setup for conducting \ac{llsi}.

\subsection{Measurement Setup}
\begin{figure}[tb]
	\centering
	\begin{subfigure}{.48\linewidth}
		\includegraphics[height=.95\linewidth, trim=190 470 0 0, clip]{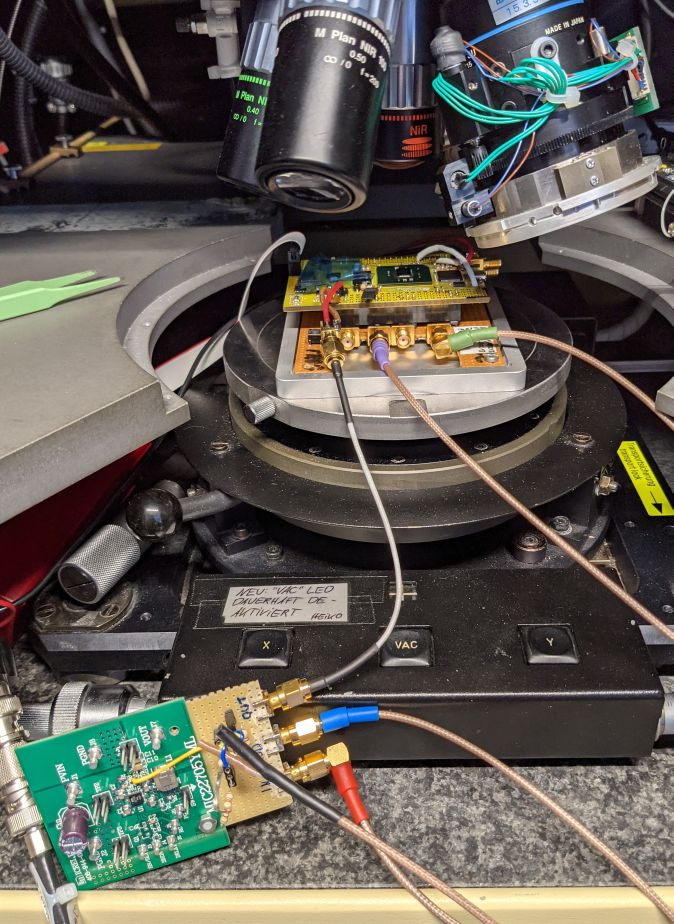}
		\caption{PHEMOS-1000\label{fig:setup_phemos_kintex}}
	\end{subfigure}
	\hfill
	\begin{subfigure}{.48\linewidth}
		\includegraphics[height=.95\linewidth]{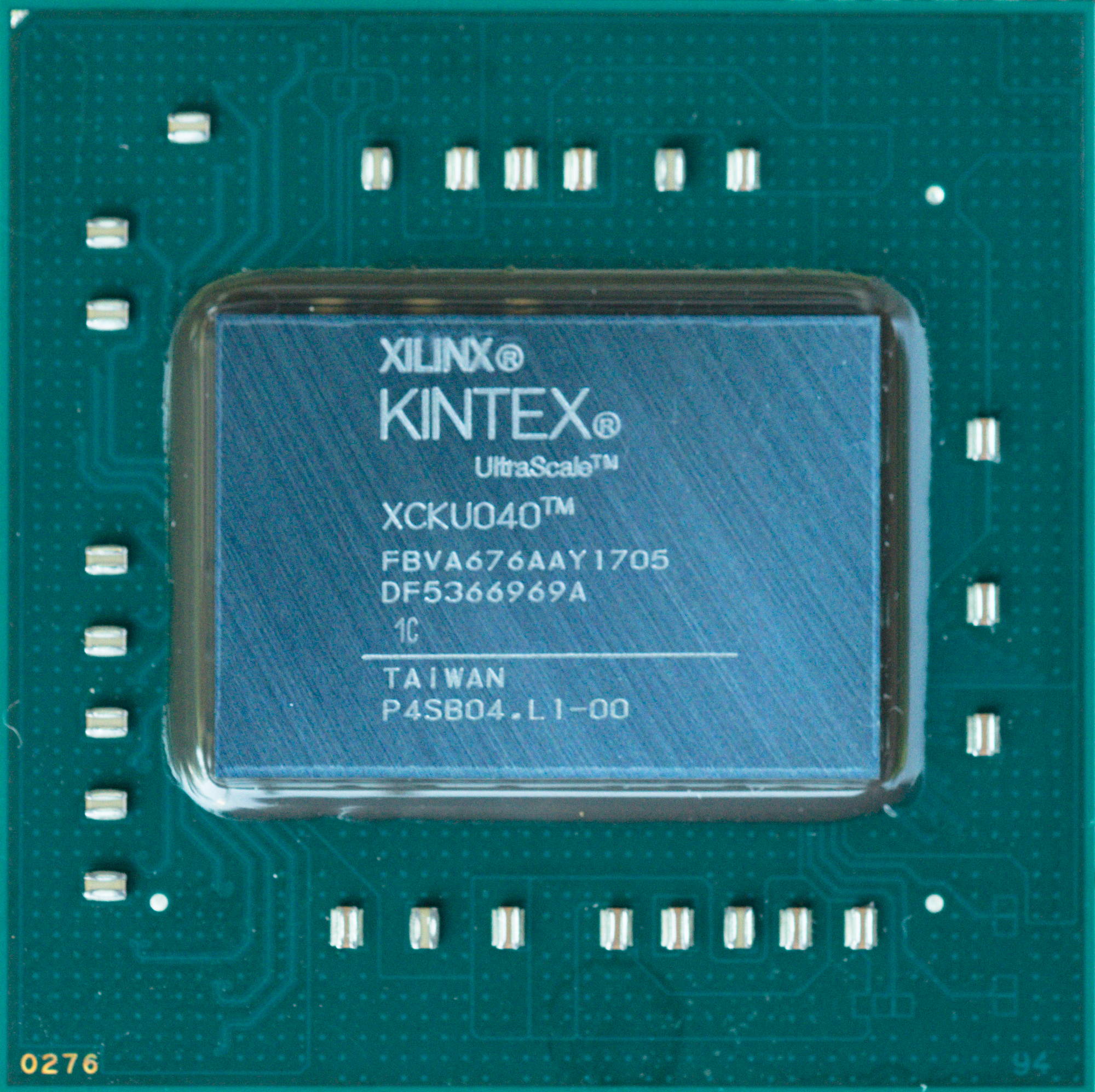}
		\caption{Flip-chip package \label{fig:setup_ultrascale_photo}}
	\end{subfigure}
	\caption{Xilinx Kintex-7 target under the PHEMOS-1000 microscope with 5$\times$ lens in use (a) and photography of the Xilinx UltraScale device (b).\label{fig:setup_DUTs}}
\end{figure}

As the setup for capturing the \ac{llsi} images, we use a Hamamatsu PHEMOS-1000 \ac{fa} microscope, see Fig.~\ref{fig:setup_phemos_kintex}, equipped with a \ac{hil} for optical probing.
The microscope offers 5$\times$, 20$\times$, and 50$\times$ lenses and an additional scanner zoom of $\times$2, $\times$4, and $\times$8.
Due to the light source's wavelength of around 1.3\,\si{\micro\meter} and the \ac{na} of our 50$\times$ lens of 0.71, the minimum beam diameter is around 1\,\si{\micro\meter}.
The step size of the galvanometric scan mirrors, however, is in the range of a few nanometers.
For \ac{eofm}/\ac{llsi} measurements, the frequency of interest $f$, the bandpass bandwidth  $\Delta f$, and the pixel dwell time $\dwelltime$ (in~\si{\milli\second\per\pixel}) can be configured in the PHEMOS software.
To achieve \ac{llsi} measurements with an acceptable noise level, it is required to modulate the power rail of interest at more than around \SI{80}{\kilo\hertz}.
In order to map the \ac{llsi} image to the exact position on the chip, an optical light reflectance image can be captured alongside the measurement.

To better evaluate the \ac{llsi} signal differences and map them to a location on the optical image, we used the ImageJ application~\cite{rueden_imagej2_2017}.
The pixel-wise subtraction of two \ac{llsi} images results in a mostly gray image with the differences displayed in white and black color.
While this already shows the differences between the images clearly, the location of the changes is not intuitively visible.
To superimpose the difference image on an optical image, we first remove noise by the "despeckle" functionality of ImageJ, and then merged the optical image and the difference image.
To improve the visibility of the differences, we have remapped the black and white spots in the raw difference image to the colors yellow and green.

\subsection{Devices Under Test}

\begin{figure}
	\centering
	\begin{subfigure}{.66\linewidth}
		\centering
		\includegraphics[width=\linewidth]{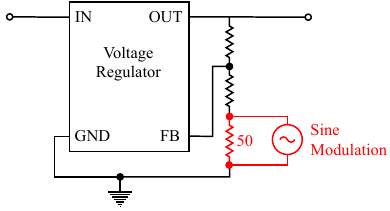}
		\caption{Schematic\label{fig:setup_modulation_schematic}}
	\end{subfigure}
	\hfill
	\begin{subfigure}{.30\linewidth}
		\centering
		\includegraphics[width=\linewidth]{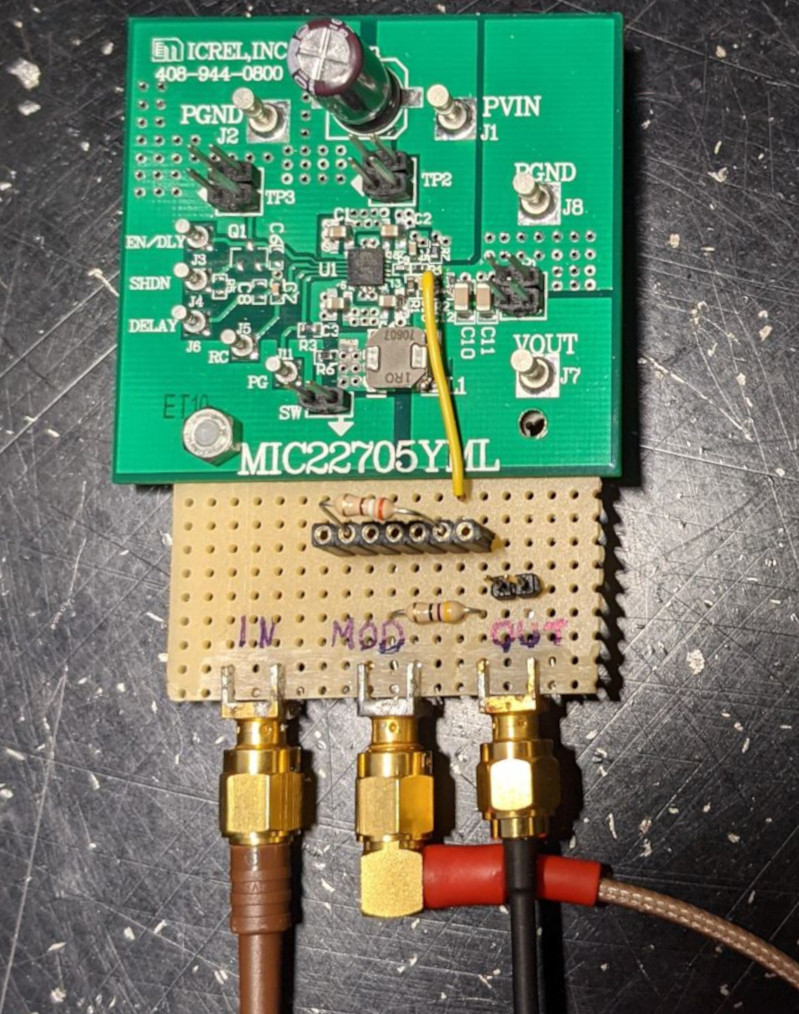}
		\caption{Modulator\label{fig:setup_modulation_mic}}
	\end{subfigure}
	\caption{LLSI modulation setup with (a) modulation regulator schematic and (b) the modified MIC22705YML-EV board.\label{fig:setup_modulation}}
\end{figure}

\begin{figure*}[tb]
	\centering
	\newlength{\dutheight}
	\setlength{\dutheight}{4.2cm}
	\newlength{\dutheightpolar}
	\setlength{\dutheightpolar}{4.2cm}
	\subcaptionbox{Xilinx Kintex-7\label{fig:setup_kintex_laser}}{
		\includegraphics{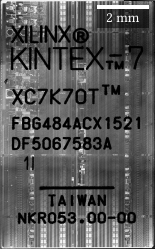}
		\includegraphics{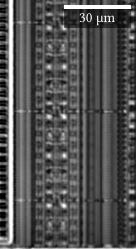}}
	\hfill
	\subcaptionbox{Xilinx UltraScale\label{fig:setup_ultrascale_laser}}{
		\includegraphics{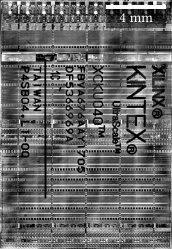}
		\includegraphics{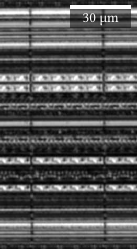}}
		\hfill
	\subcaptionbox{Microchip PolarFire SoC\label{fig:setup_polarfire_laser}}{
		\includegraphics{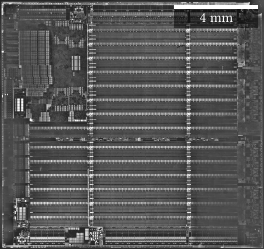}
		\includegraphics{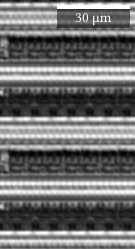}}
	\caption{Laser reflection images of the \acsp{dut}: entire chip (left) and zoom into the logic fabric (right). \label{fig:setup_DUTs_laser}}
\end{figure*}

\subsubsection{Xilinx Kintex-7 FPGA\label{sec:setup_kintex}}
As SRAM-based \ac{fpga}, we chose the Xilinx Kintex-7 XC7K70T, manufactured in a \SI{28}{\nano\meter} technology.
The chip is available in a \ac{bga} bare-die flip-chip package on a Numato Systems Skoll development board. %
The FPGA can be programmed using the Xilinx Vivado \ac{ide}.
In the Kintex-7 architecture~\cite{xilinxin_series_2016}, the logic fabric is comprised of \acp{clb}, which consist of two so-called logic slices, and have a switch matrix for connecting to the global routing matrix.
One slice consists of four 6-input \acp{lut} (which can be configured as two 5-input \acp{lut} with separate outputs each), eight \acp{ff}, as well as \acp{mux} and arithmetic carry logic.
While the slice naming uses~X and~Y coordinates (e.g., SLICE\_X0Y0), the \acp{lut} inside one slice are named from A5LUT/A6LUT to D5LUT/D6LUT, and the corresponding \acp{ff} from AFF/A5FF to DFF/D5FF.
Next to the logic slices (2/3 of all slices), there are also memory slices usable as distributed RAM or shift registers.

\begin{figure}
	\centering
	\includegraphics[width=\linewidth, trim=0 100 0 50, clip]{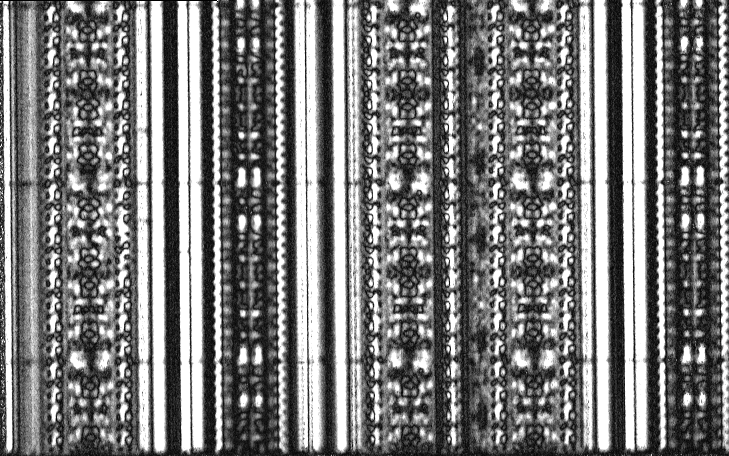}
	\caption{LLSI raw image from the logic fabric on the Kintex-7 \ac{fpga}. 50$\times$\,($\times$2) zoom, $\dwelltime = \SI{2.1}{\milli\second/\pixel}$, $\Delta f = \SI{300}{\Hz}$.}
	\label{fig:raw_llsi}
\end{figure}

To prepare the device for LLSI measurements, we disabled the onboard voltage regulator for VCC.
Then, we soldered an SMA connector to the voltage rail for supplying the voltage externally via a power supply that can be modulated.
For this purpose, we modified a MIC22705YML-EV voltage regulator evaluation board by replacing the resistor between the feedback pin and GND with a resistor to set the correct output voltage, in series with a 50\,$\Omega$ resistor, see Fig.~\ref{fig:setup_modulation}.
In parallel to the latter, we connected a Keithley 3390 laboratory waveform generator to generate a sine wave.
The regulator's output then provides a sine wave with a frequency of up to \SI{300}{\kilo\hertz} and a DC offset of the rated value for VCC of 1\,\si{\volt} with a sufficient current drive strength.
For higher frequencies, the regulator would stop functioning as intended.
However, already when trying to modulate the \ac{dut}'s voltage rail at low frequencies of a few \si{\kilo\hertz}, no significant modulation can be measured on the \ac{pcb}'s voltage rail.
The reason for that is the existence of large decoupling capacitors, smoothing undesired peaks and fluctuations of the supply voltage.
We desoldered all decoupling capacitors connected to VCC of \SI{0.1}{\micro\farad} and larger using a hot air station to achieve a sufficiently high modulation amplitude.
As a result, we could achieve a peak-to-peak modulation between \SI{150}{\mV} and \SI{200}{\mV} around the VCC offset of \SI{1}{\volt} at a frequency $f$ of \SI{80}{\kilo\hertz}.

Fig.~\ref{fig:setup_kintex_laser} shows optical (light reflectance) images of the entire chip and a section of the logic fabric.
A raw \ac{llsi} image from the Kintex-7 logic fabric indicates that the modulation of VCC influences the light reflection almost everywhere, see Fig.~\ref{fig:raw_llsi}.

\subsubsection{Xilinx UltraScale FPGA\label{sec:setup_ultrascale}}
As a second SRAM-based \ac{fpga}, we chose the Xilinx UltraScale XCKU040, manufactured in a \SI{20}{\nano\meter} technology.
The chip is available in a flip-chip bare-die package, see Fig.~\ref{fig:setup_ultrascale_photo}, on an AVNET development board (model AES-KU040-DB-G).
Similar to the Kintex-7 architecture (Section \ref{sec:setup_kintex}), the UltraScale logic fabric is comprised of \acp{clb}.
Each \ac{clb} contains one slice providing eight 6-input \acp{lut} (which can also be configured as two 5-input \acp{lut} with separate outputs), sixteen \acp{ff}, as well as \acp{mux} and arithmetic carry logic.
The slices are named using X and Y coordinates, whereas the \acp{lut} and \acp{ff} are named with capital letters (A5LUT/A6LUT to H5LUT/H6LUT and AFF/AFF2 to HFF/HFF2).
Next to the logic slices, there are memory slices that can be used as distributed RAM or shift registers.
Fig.~\ref{fig:setup_ultrascale_laser} shows optical images of the entire chip and a section of the logic fabric.

To modulate the voltage rail of the UltraScale target, we used the same external modulation circuit as for the Kintex-7 (see Fig.~\ref{fig:setup_modulation}).
First, we disabled the onboard voltage regulator for VCC (\SI{0.95}{\volt}) by desoldering the coil at the regulator's output.
Then, we soldered an SMA connector to the corresponding pad for supplying VCC externally.
Furthermore, we desoldered all decoupling capacitors connected to VCC of \SI{0.1}{\micro\farad} and larger from the \ac{pcb} for being able to modulate the voltage rail at a sufficiently high frequency.
For the experiments, we used a peak-to-peak modulation of around \SI{150}{\mV} at a frequency $f$ of \SI{80}{\kilo\hertz} with a VCC offset of \SI{0.95}{\volt}.

\subsubsection{Microchip PolarFire SoC FPGA\label{sec:setup_polarfire}}
As flash-based FPGA, we chose the Microchip PolarFire SoC MPFS250T-FCVG484EES, manufactured in a 28\,\si{\nano\meter} technology.
The configuration is stored in distributed flash cells manufactured in Microchip's SONOS technology~\cite{microsemi_white_2017}, consisting of two floating-gate transistors.
The chip is available on the PolarFire SoC FPGA Icicle Kit %
in a \ac{bga} flip-chip package with a lid.
After cooling down the device in a typical household freezer, we could pry off the lid using a knife to access the chip backside.
The FPGA can be programmed using the Microsemi Libero \ac{ide}.
In the PolarFire architecture~\cite{microchip_ug0680_2021}, the logic fabric is comprised of arrays of \acp{lc} that are connected by \ac{il}.
Each \ac{lc} consists of 12 \acp{le}, whereas each \ac{le} contains a 4-input \ac{lut}, a \ac{ff}, and a \ac{mux}.
Next to a connection to the \ac{il}, the individual \acp{le} inside one \ac{lc} are connected by a carry chain.
Next to the \acp{lc}, there are other blocks, such as dedicated math and memory blocks, connected via the \ac{il}.

We could use the onboard MIC22705YML voltage regulator for modulating VDD %
of this target.
Via a jumper, the resistor in the feedback path can be changed to create a \SI{1.0}{\volt} or \SI{1.05}{\volt} supply voltage.
By removing the jumper and connecting our own resistors, we could create the same modulation capabilities as shown in Fig.~\ref{fig:setup_modulation_schematic}.
To increase the \ac{llsi} signal's amplitude, we desoldered all decoupling capacitors connected to VDD of \SI{0.1}{\micro\farad} and larger from the \ac{pcb}.
We used a peak-to-peak modulation of approximately 170\,\si{\milli\volt} around the VDD offset of 1\,\si{\volt}.
A modulation frequency $f$ of \SI{83.5}{\kilo\hertz} led to the highest \ac{llsi} signal amplitude.
Note that the SONOS cells are not supplied by VDD but VDD25, which is supplied by a \SI{2.5}{\volt} regulator.
To modulate the VDD25 voltage, we soldered a jumper to disable the onboard regulator and added an SMA connector to supply VDD25 via our external modulator circuit.
However, as we could not detect any benefit over modulating VDD, we only used the VDD modulation for the experiments presented in this paper. 
Fig.~\ref{fig:setup_polarfire_laser} shows optical images of the entire chip and a part of the logic fabric.

\section{Results \label{sec:results}}

\subsection{Detecting Changes in the Logic Fabric\label{sec:results:logic}}

To investigate the capabilities of \ac{llsi} for detecting changes in the logic fabric configuration, we first tried to detect small changes within one logic element, i.e., changes in the \ac{lut} configurations and \ac{ff} logic states.
Although the number of different configurations is high, we aimed at creating a good coverage of detectable changes.

\subsubsection{SRAM-based (Kintex-7)}

\begin{figure}
	\centering
	\newlength{\kintexllsiwidth}
	\setlength{\kintexllsiwidth}{.28\linewidth}
	\begin{subfigure}{\linewidth}
		\centering
		\includegraphics{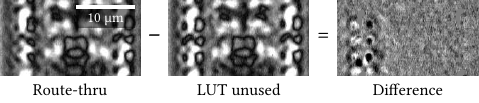}
		\captionsetup{skip=1pt}
		\caption{LUT used (route-thru) vs. unused\label{fig:results_SRAM_LUTused}}
	\end{subfigure}
	\begin{subfigure}{\linewidth}
		\centering
		\vspace{2mm}
		\includegraphics{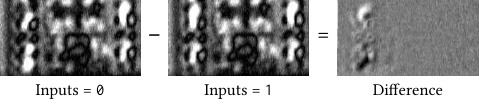}
		\captionsetup{skip=1pt}
		\caption{5-input LUT with all inputs set to one value\label{fig:results_SRAM_LUTinp}}
	\end{subfigure}
	\begin{subfigure}{\linewidth}
		\centering
		\vspace{2mm}
		\includegraphics{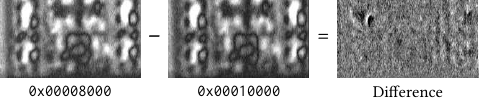}
		\captionsetup{skip=1pt}
		\caption{1-bit LUT configuration (\texttt{INIT}) value change \label{fig:results_SRAM_config_small}}
	\end{subfigure}
	\begin{subfigure}{\linewidth}
		\centering
		\vspace{2mm}
		\includegraphics{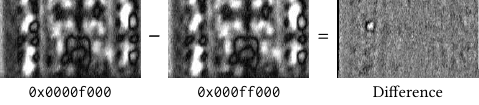}
		\captionsetup{skip=1pt}
		\caption{4-bit LUT configuration (\texttt{INIT}) value change \label{fig:results_SRAM_config_big}}
	\end{subfigure}
	\caption{Kintex-7 LLSI results for different lookup-table configurations.
		50$\times$\,($\times$4) zoom, $\dwelltime = \SI{3.3}{\milli\second/\pixel}$, $\Delta f = \SI{100}{\Hz}$.\label{fig:results_SRAM}}
\end{figure}

\mysubsubsubsection{LUT used vs. unused}
We compared implementations where once the \ac{lut} is unused and once a route-thru \ac{lut} is implemented.
We assumed a route-thru \ac{lut} to be the configuration with minimal differences compared to the unused \ac{lut}, as the input of the \ac{lut} is directly routed to the output of the SLICE.
Nevertheless, the differences can be clearly identified, see Fig.~\ref{fig:results_SRAM_LUTused}.

\mysubsubsubsection{LUT inputs 0 vs. 1}
When changing the values of \ac{lut} inputs, which originate from the output of another \ac{lut} or a \ac{ff}, the change is clearly visible as well, see Fig.~\ref{fig:results_SRAM_LUTinp}.
As could be expected, we observed fewer changes if fewer input values are changed.
Still, we could detect changes also if only one input value is changed.

\mysubsubsubsection{LUT configuration value changes}
The smallest possible change we could imagine is the manipulation of single bits in the \ac{lut} configuration.
We observed that the number of bits changed in the \ac{lut} configuration \texttt{INIT} value does not necessarily determine how significant the difference in the \ac{llsi} response is, see Figs.~\ref{fig:results_SRAM_config_small} and~\ref{fig:results_SRAM_config_big}.
We assume that not the SRAM cell holding the configuration produces the LLSI signature, but the actual multiplexers and pass transistors.
If a configuration change causes -- due to the applied \ac{lut} inputs -- more multiplexers to change their states (cf. Fig.\ref{fig:background_le}), there will be a bigger difference between the LLSI images.

\begin{figure}
	\centering
	\begin{subfigure}{.44\linewidth}
		\centering
		\includegraphics{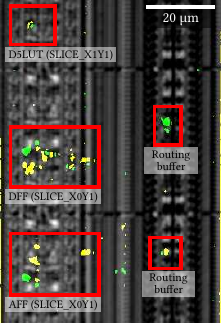}
	\caption{LLSI difference}
\end{subfigure}
\hspace{2mm}
\begin{subfigure}{.48\linewidth}
	\centering
	\includegraphics[width=\linewidth, trim=0 -.6cm 0 0, clip]{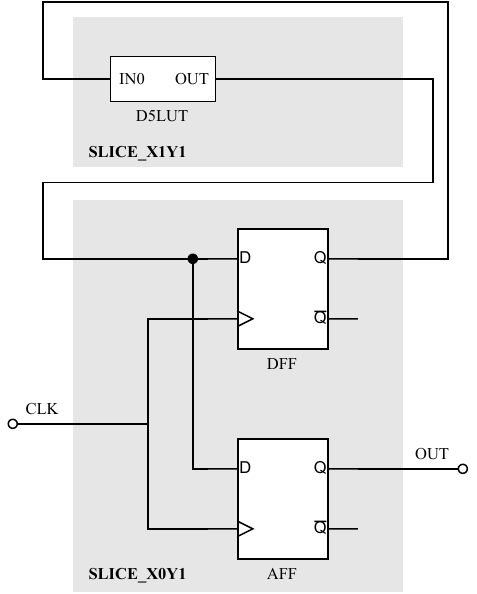}
	\caption{Logic schematic}
\end{subfigure}
\caption{Kintex-7 LLSI difference superimposed over an optical image for \ac{ff} values \texttt{0} vs. \texttt{1} with CLB inputs and outputs connected.
	Yellow and green colors correspond to the black and white spots in the raw difference image.
	50$\times$\,($\times$2) zoom, $\dwelltime = \SI{2.1}{\milli\second/\pixel}$, $\Delta f = \SI{300}{\Hz}$.\label{fig:results_SRAM_register}}
\end{figure}

\mysubsubsubsection{FF value 0 vs. 1}
Finally, we designed a bit more complex design, which contains two \acp{ff} and one \ac{lut} residing in different logic slices, see Fig.~\ref{fig:results_SRAM_register}.
We have subtracted the LLSI images of two consecutive clock cycles.
While the difference for the \ac{lut} is concentrated in a single small area, there are many different spots for the \acp{ff}.
This might be explained by the fact that the input buffers, the actual memory cell, the output buffers, and the clock buffers have changed their values by advancing a clock cycle as well.
Interestingly, although the two registers were instantiated in exactly the same way in the \ac{ide}, different changes can be observed between them.
This might be caused by the different output configurations of the \acp{ff} or an asymmetric \ac{asic} design of the \ac{clb}.
For instance, the clock buffers or some intra-\ac{clb} routing capabilities, which are invisible in the \ac{ide} for the designer, might reside close to DFF.
Finally, we could observe differences in the (assumed-to-be) routing areas, supposedly interconnecting the two slices X0Y1 and X1Y1.

\subsubsection{SRAM-based (UltraScale)}
To investigate if similar results can be achieved on a \ac{dut} manufactured in a smaller technology, we conducted the same experiments on the UltraScale FPGA.

\begin{figure}
	\centering
	\newlength{\ultrascalellsiwidth}
	\setlength{\ultrascalellsiwidth}{.28\linewidth}
	\newcommand{\ultrascalellsiang}{90}
	\begin{subfigure}{\linewidth}
		\centering
		\includegraphics{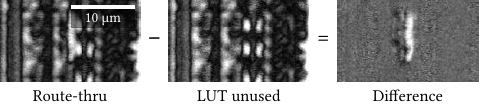}
		\captionsetup{skip=1pt}
		\caption{LUT used (route-thru) vs. unused\label{fig:results_LUT_us_LUTused}}
	\end{subfigure}
	\begin{subfigure}{\linewidth}
		\centering
		\vspace{2mm}
		\includegraphics{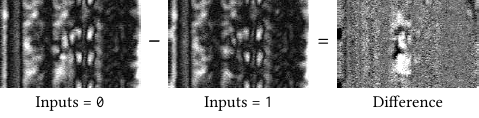}
		\captionsetup{skip=1pt}
		\caption{5-input LUT with all inputs set to one value\label{fig:results_LUT_us_LUTinp}}
	\end{subfigure}
	\begin{subfigure}{\linewidth}
		\centering
		\vspace{2mm}
		\includegraphics{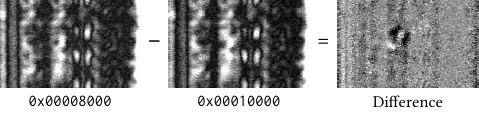}
		\captionsetup{skip=1pt}
		\caption{1-bit LUT configuration (\texttt{INIT}) value change \label{fig:results_LUT_us_config_small}}
	\end{subfigure}
	\begin{subfigure}{\linewidth}
		\centering
		\vspace{2mm}
		\includegraphics{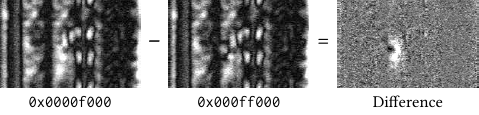}
		\captionsetup{skip=1pt}
		\caption{4-bit LUT configuration (\texttt{INIT}) value change \label{fig:results_LUT_us_config_big}}
	\end{subfigure}
	\caption{UltraScale LLSI results for different lookup-table configurations.
		50$\times$\,($\times$4) zoom, $\dwelltime = \SI{2.1}{\milli\second/\pixel}$, $\Delta f = \SI{300}{\Hz}$.\label{fig:results_LUT_us}}
\end{figure}

\mysubsubsubsection{LUT used vs. unused}
Although the technology node size of the UltraScale series is around 28\% smaller than of the Kintex-7 series, the difference between a route-thru \ac{lut} and a completely unused \ac{lut} is clearly visible, see Fig.~\ref{fig:results_LUT_us_LUTused}.
Due to the technology size reduction, the affected area is smaller but can still be resolved using our optical setup.
Furthermore, the difference image looks more blurry than for the Kintex-7 FPGA.
One explanation for this might be the lower modulation amplitude achievable on the UltraScale board.

\mysubsubsubsection{LUT inputs 0 vs. 1}
Flipping the \ac{lut}'s inputs values can be detected reliably as well, see Fig.~\ref{fig:results_LUT_us_LUTinp}.
Interestingly, the affected area seems to be as large as in the previous experiment on used vs. unused \ac{lut}.
The reason might be that we can not control the routing of signals and which values are applied to unused inputs.

\mysubsubsubsection{LUT configuration value changes}
We could clearly detect the same \ac{lut} configuration changes that we could detect on the Kintex-7, see Figs.~\ref{fig:results_LUT_us_config_small} and~\ref{fig:results_LUT_us_config_big}.
For this target, the affected area neither reflects the number of bits changed in the configuration.
This observation supports the hypothesis that the \ac{lut}'s multiplexers and not the memory cells for the configuration contribute most to the \ac{llsi} signal.

\begin{figure}
	\centering
	\begin{subfigure}{.46\linewidth}
		\centering
		\includegraphics{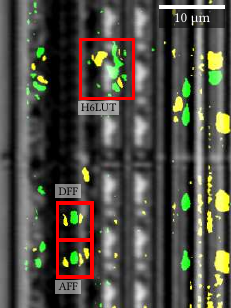}
	\caption{LLSI difference}
\end{subfigure}
\begin{subfigure}{.52\linewidth}
	\centering
	\includegraphics[width=\linewidth]{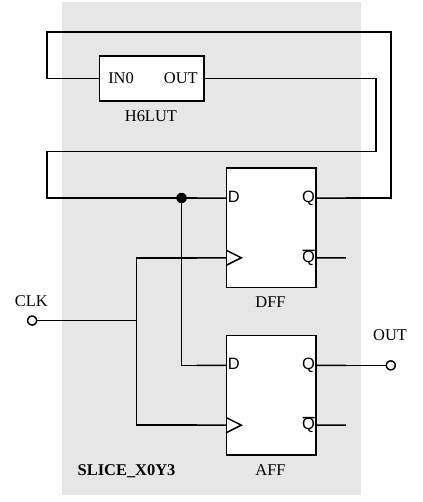}
	\caption{Logic schematic}
\end{subfigure}
\caption{UltraScale LLSI difference superimposed over an optical image for different \ac{ff} values and \ac{lut} inputs. 
	50$\times$\,($\times$4) zoom, $\dwelltime = \SI{2.1}{\milli\second/\pixel}$, $\Delta f = \SI{300}{\Hz}$.\label{fig:results_Ultrascale_register}}
\end{figure}

\mysubsubsubsection{FF value 0 vs. 1}
When investigating an entire \ac{clb} with one \ac{lut} and two \acp{ff} in use, multiple areas with differences in the \ac{llsi} image can be observed, see Fig.~\ref{fig:results_Ultrascale_register}.
Again, we subtracted the \ac{llsi} images of two consecutive clock cycles.
From the knowledge gained in the previous experiments, we could identify the changes in the \ac{lut} and map two areas with similar changes to the two \acp{ff}.
Despite these distinctly allocable changes, many other areas with clear differences appear in the image.
These changes seem to belong to the \ac{clb}'s \acp{mux} (left of the \acp{lut} and \acp{ff}) and routing resources, such as buffers (right side of the image).
However, since the chip's layout is unknown, these assumptions can not be verified further.

\subsubsection{Flash-based (PolarFire SoC)}
To investigate whether configuration changes can also be detected on the flash-based FPGA, we conducted similar experiments on the PolarFire SoC FPGA.

\begin{figure}
\centering
\newlength{\polarllsiwidth}
\setlength{\polarllsiwidth}{.18\linewidth}
\begin{subfigure}{\linewidth}
	\centering
	\includegraphics{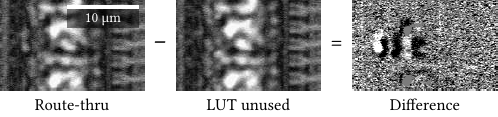}
	\captionsetup{skip=1pt}
	\caption{LUT used (route-thru) vs. unused\label{fig:results_Flash_LUTused}}
\end{subfigure}
\begin{subfigure}{\linewidth}
	\centering
	\vspace{2mm}
	\includegraphics{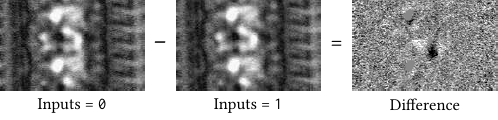}
	\captionsetup{skip=1pt}
	\caption{4-input LUT with all inputs set to one value\label{fig:results_Flash_LUTinp}}
\end{subfigure}
\begin{subfigure}{\linewidth}
	\centering
	\vspace{2mm}
	\includegraphics{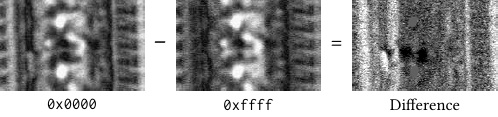}
	\captionsetup{skip=1pt}
	\caption{Large LUT configuration (\texttt{INIT}) value change\label{fig:results_Flash_config_0000_1111}}
\end{subfigure}
\begin{subfigure}{\linewidth}
	\centering
	\vspace{2mm}
	\includegraphics{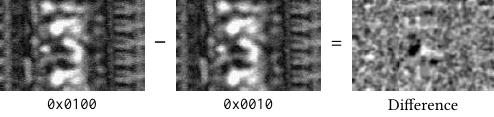}
	\captionsetup{skip=1pt}
	\caption{Small LUT configuration (\texttt{INIT}) value change\label{fig:results_Flash_config_input_0}}
\end{subfigure}
\caption{PolarFire SoC LLSI results.
	Images rotated by 90 degrees, 50$\times$\,($\times$4) zoom, $\dwelltime = \SI{3.3}{\milli\second/\pixel}$, $\Delta f = \SI{100}{\Hz}$.\label{fig:results_Flash}}
\end{figure}

\mysubsubsubsection{LUT used vs. unused}
For this target, we compared the configuration for a route-thru \ac{lut} with an unused \ac{lut} as well, see Fig.~\ref{fig:results_Flash_LUTused}.
The LLSI responses show a clear difference, although the corresponding area is smaller than on the Xilinx FPGAs.
The reason might be that the \acp{lut} on Kintex-7 and UltraScale have up to 6 inputs, while they only have 4 inputs on PolarFire, resulting in a significant difference in the number of contained \acp{mux}.

\mysubsubsubsection{LUT inputs 0 vs. 1}
The area of differences when only the \ac{lut} inputs change are smaller than the differences between a used and unused \ac{lut} -- as can be expected, see Fig.~\ref{fig:results_Flash_LUTinp}.

\mysubsubsubsection{LUT configuration value changes}
Changes in the \ac{lut} configurations can be detected as well.
For a large change in the configuration, i.e., by flipping all bits, the change with the largest area is visible, see Fig.~\ref{fig:results_Flash_config_0000_1111}.
As for the other FPGAs, the reason might be the different number of \acp{mux} affected by the configuration change, under the assumption that the inputs of the \ac{lut} stay constant.
For a 2-bit change in the \texttt{INIT} value, a smaller difference is visible, see Fig.~\ref{fig:results_Flash_config_input_0}.
Moreover, we observed that when all \ac{lut} inputs are set to \texttt{0}, the difference for changed \texttt{INIT} values is larger than when all inputs are set to \texttt{1}.
Since in our experiment the output of the \ac{lut} was not changed by applying the different inputs (due to the configured \texttt{INIT} value), we suppose that a different number of multiplexers changed their states depending on the \ac{lut} inputs.

\begin{figure}
\centering
\begin{subfigure}{.42\linewidth}
	\centering
	\includegraphics{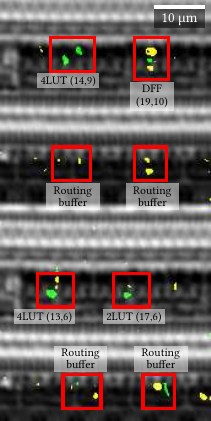}
\caption{LLSI difference}
\end{subfigure}
\begin{subfigure}{.53\linewidth}
\centering
\vspace{14mm}
\includegraphics[width=\linewidth, trim=0 -1.5cm 0 0, clip]{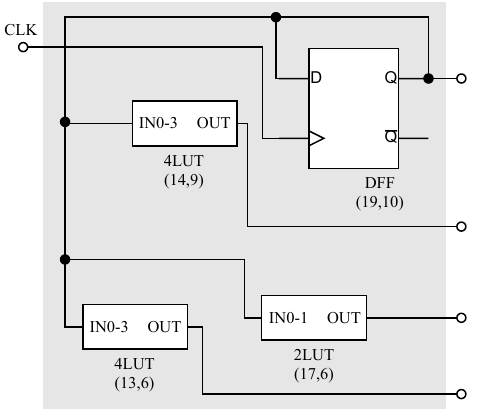}
\vspace{1mm}
\caption{Logic schematic}
\end{subfigure}
\caption{PolarFire SoC LLSI difference superimposed over an optical image for different \ac{ff} values and \ac{lut} inputs. 
50$\times$\,($\times$2) zoom, $\dwelltime = \SI{3.3}{\milli\second/\pixel}$, $\Delta f = \SI{100}{\Hz}$.\label{fig:results_Flash_register}}
\end{figure}

\mysubsubsubsection{FF value 0 vs. 1}
Similar to the experiments on the SRAM-based FPGAs, we created snapshots of a larger area of the logic fabric, on the one hand, to observe the LLSI response differences for a \ac{ff}, and on the other hand, to learn about the detectability of buffers and routing transistors. Fig.~\ref{fig:results_Flash_register} shows the difference of two LLSI responses captured in two consecutive clock cycles.
The state change of the \ac{ff} is clearly visible on the top right of the image.
The three \acp{lut} receive the output of the \ac{ff} as inputs, and therefore, their responses differ, too.
Differences can also be observed in between the rows of logic elements.
These areas presumably belong to the routing logic, thus containing data and clock buffers.

\subsection{Detecting Changes in Routing}
\begin{figure}
\centering
\begin{subfigure}{\linewidth}
\centering
\includegraphics[width=\linewidth, trim=0 2.5cm 0 1.5cm, clip]{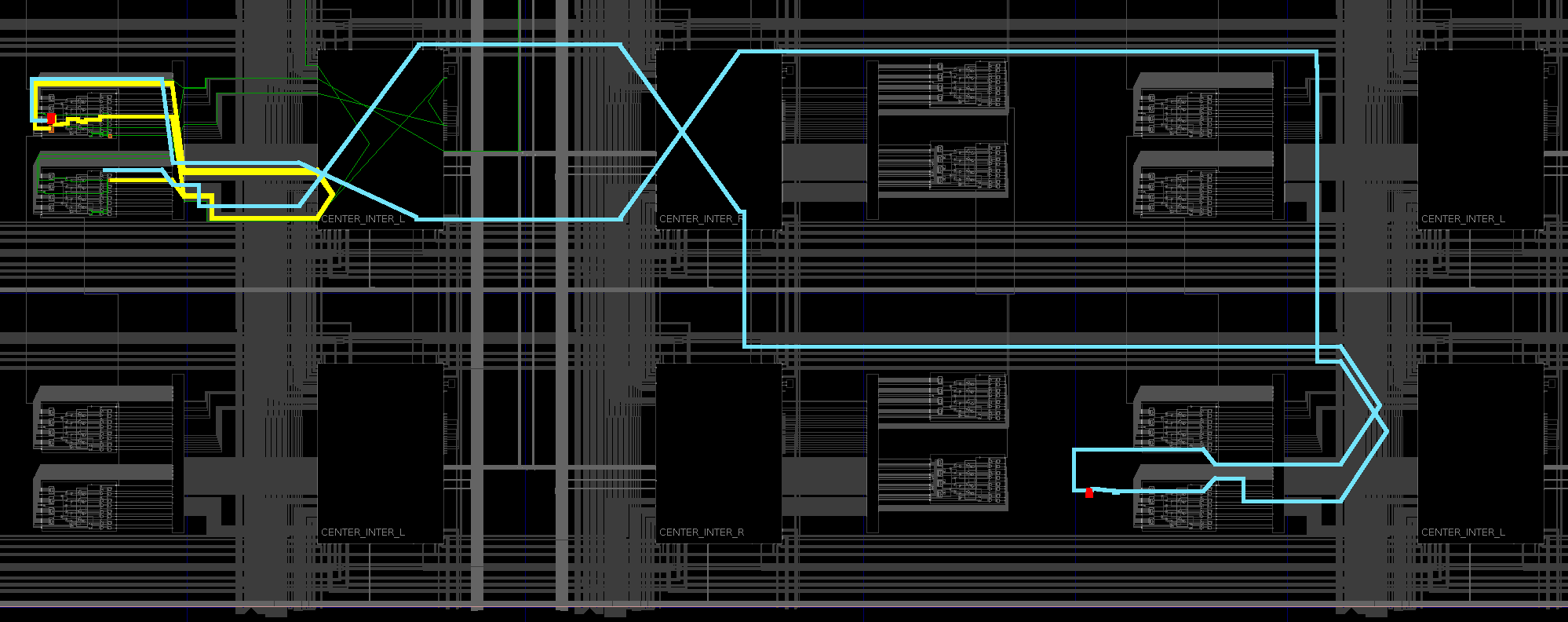}
\caption{Placement schematic of design with LUT in SLICE\_X1Y1 (yellow) and in SLICE\_X4Y0 (blue)\label{fig:routing_kintex_orig}}
\end{subfigure}
\begin{subfigure}{\linewidth}
\centering
\vspace{2mm}
\includegraphics{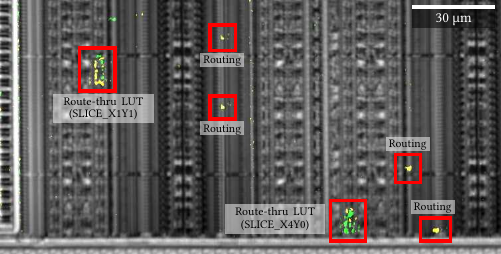}
\caption{LLSI difference\label{fig:routing_kintex_LLSI}}
\end{subfigure}
\caption{Difference in routing configuration on Kintex-7 when moving a< route-thru LUT from SLICE X1Y1 to  X4Y0 while keeping the signal source and destination in SLICE X1Y1 and X0Y1.
50$\times$\,($\times$2) zoom, $\dwelltime = \SI{2.1}{\milli\second/\pixel}$, $\Delta f = \SI{300}{\Hz}$.\label{fig:routing_kintex}}
\end{figure}

The authors of \cite{ender_first_2017} propose malicious modifications in the signal runtime on the \ac{fpga} by using either route-thru \acp{lut} or manipulating the routing to take longer paths.
We have already shown that the insertion of route-thru \acp{lut} can be detected; see Section~\ref{sec:results:logic}.
To test the capability of our approach to detect changes in the routing, we created a design for the Kintex-7 FPGA that contains one route-thru \ac{lut}, whose location we change between two measurements.
Thereby, the signal is forced to be routed differently. 
For the first snapshot, the \ac{lut} is placed in SLICE\_X1Y1, while for the second snapshot, it is placed in SLICE\_X4Y0, see Fig.~\ref{fig:routing_kintex_orig}. The signal source and sink are kept at the same location (in SLICE\_X0Y1 and X1Y1). Fig.~\ref{fig:routing_kintex_LLSI} clearly shows not only the differences in the \ac{llsi} response for the changed \ac{lut} placement but also for the routing logic.
Consequently, one can also detect changes in signal routing with our approach.

\subsection{Trojan Benchmarks}
The previous results have already shown that small changes, down to single bit changes in the \ac{lut} configuration and small changes in the routing configuration, can be detected using our method.
Therefore, we have demonstrated that \ac{llsi} can detect the malicious modifications proposed in \cite{ender_first_2017} introducing changes in the signal path delays.
To demonstrate that we can also detect other \acp{ht} proposed in the literature, we exemplarily implemented \ac{ht} benchmarks generated using the TRIT framework \cite{cruz_automated_2018}, which can be found on TrustHub~\cite{shakya_benchmarking_2017}.
We implemented two benchmarks on the Kintex-7 \ac{dut}, one consisting only of combinatorial \ac{ht} logic (from TRIT-TC) and one also containing sequential logic (from TRIT-TS).
All provided benchmarks generated using TRIT introduce additional logic gates and/or \acp{ff}.
We fixed the location and routing placement of all logic components and the routing that does not belong to the \ac{ht} trigger or payload to keep the changes of the implementation minimal.

\subsubsection{Combinatorial Trojan}
\begin{figure}[tb]
\centering
\begin{subfigure}{.49\linewidth}
\centering
\hfill
\includegraphics{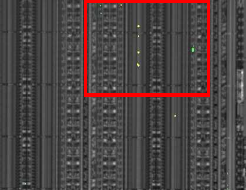}
\caption{50$\times$ zoom\label{fig:results_troj_seq_SRAM_1x}}
\end{subfigure}
\hfill
\begin{subfigure}{.49\linewidth}
\centering
\includegraphics[width=\linewidth]{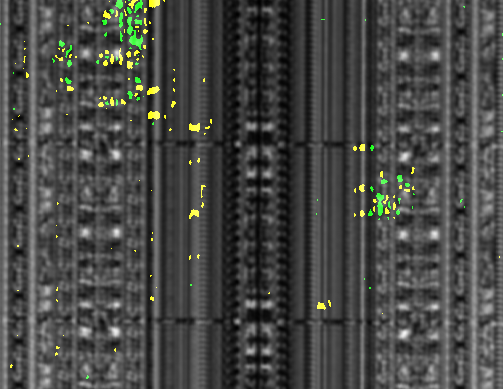}
\caption{50$\times$\,($\times$2) zoom\label{fig:results_troj_seq_SRAM_2x}}
\end{subfigure}
\hfill
\vspace{-2mm}
\caption{Combinatorial Trojan benchmark (c2670\_T071) section on Kintex-7. (a) $\dwelltime = \SI{5}{\milli\second/\pixel}$, (b) $\dwelltime = \SI{3.3}{\milli\second/\pixel}$, $\Delta f = \SI{100}{\Hz}$. \label{fig:results_troj_seq_SRAM}}
\end{figure}

The \emph{c2670\_T071} \ac{ht} benchmark introduces six additional logic gates.
Fig.~\ref{fig:results_troj_seq_SRAM} only shows a part of the logic fabric area consumed by the implementation.
However, already in this section of the design, clear differences can be observed.
As can be seen, zooming into an area with suspicious differences can highlight the changes more clearly.

\subsubsection{Sequential Trojan}
\begin{figure}[tb]
\centering
\begin{subfigure}[t]{.49\linewidth}
\centering
\hfill
\includegraphics[width=\linewidth]{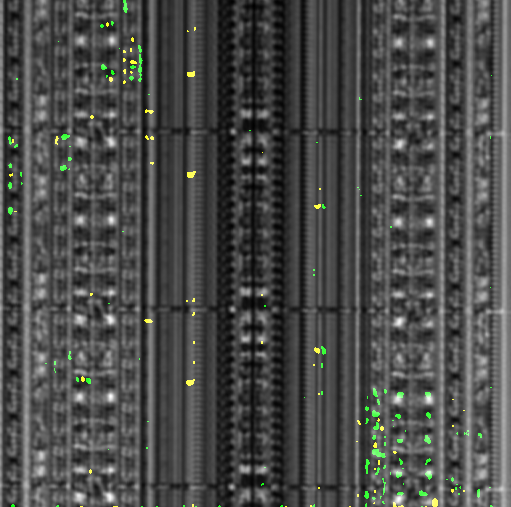}
\caption{Free vs. Trojan\label{fig:results_troj_comb_SRAM_freeVsTroj}}
\end{subfigure}
\hfill
\begin{subfigure}[t]{.49\linewidth}
\centering
\includegraphics[width=\linewidth]{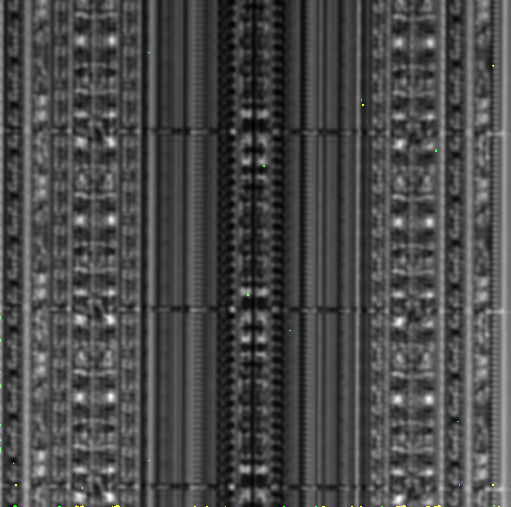}
\caption{Free vs. Free, different runs\label{fig:results_troj_comb_SRAM_freeVsFree}}
\end{subfigure}
\hfill
\vspace{-1mm}
\caption{Sequential Trojan benchmark (s1423\_T607) section on Kintex-7. 50$\times$\,($\times$2) zoom, $\dwelltime = \SI{3.3}{\milli\second/\pixel}$, $\Delta f = \SI{100}{\Hz}$.\label{fig:results_troj_comb_SRAM}}
\end{figure}

Next to combinatorial gates, the \emph{s1423\_T607} benchmark contains a counter with 15 states implemented using \acp{ff}.
Fig.~\ref{fig:results_troj_comb_SRAM_freeVsTroj} indicates that many changes can be detected both in the \acp{clb} and routing areas.
As expected, when capturing two \ac{llsi} images of the same area from the Trojan-free design, no clear differences can be observed, see Fig.~\ref{fig:results_troj_comb_SRAM_freeVsFree}.
This proves that the previously observed differences are not only caused by noisy measurements.

\section{Discussion \label{sec:discussion}}
In this section, we first discuss further research directions continuing our approach.
Subsequently, we talk about the applicability of our approach and discuss potential limitations.

\subsection{Further Research Directions\label{sec:discussion:future}}

\subsubsection{Application to ASICs\label{sec:discussion_asics}}
Regarding the applicability of our approach to \ac{asic} implementations, a few things have to be kept in mind.
Generally, it should be possible to detect the locations of all transistors and then overlay the layout file.
In this way, irregularities and deviations from the intended designs can be detected, even without having a golden chip.
One drawback is that modifications that only affect the metal layers can not be detected if the changes do not manifest in the light reflection.
However, we think that detecting analog \acp{ht}, such as capacitor-based and dopant-level Trojans, should be possible using \ac{llsi}.
Since these \acp{ht} use analog properties of the chip and are pre-silicon modifications, we could not investigate them.
However, in the following, we explain why our approach should be able to detect such \acp{ht}.

\mysubsubsubsection{Detecting capacitor-based Trojans}
Results from \cite{niu_laser_2014} indicate that decoupling capacitors can be imaged using \ac{llsi}. %
Since these capacitors are connected between VCC and GND, the power supply modulation will modulate the electric field and charge density of the capacitor, which influences the light reflection.
Therefore, \ac{llsi} might also be applicable to detect \acp{ht} that only introduce changes in the capacitance to create a stealthy trigger mechanism (e.g., A2 Trojans~\cite{yang_a2_2016}).

\mysubsubsubsection{Detecting dopant-level Trojans}
The investigations in \cite{kindereit_quantitative_2007} and \cite{kindereit_investigation_2009} show that the light reflection for optical probing depends on the doping level of the silicon.
Therefore, malicious modifications in the doping concentration to alter the functionality of logic gates~\cite{becker_stealthy_2013} might be detectable using \ac{llsi}.

\subsubsection{Reverse-Engineering the FPGA Configuration}
As already shown in this work, the configuration of the \ac{fpga} logic fabric is contained in the \ac{llsi} snapshots.
Although the resolution seems to be insufficient to extract the exact configurations manually, machine learning approaches might be able to solve that task.
The advantages of employing deep learning techniques have already been demonstrated in~\cite{krachenfel_automatic_2021} for data extraction from dedicated on-chip memories.
Such configuration extraction can also facilitate the structural and functional reverse engineering of bitstreams in proprietary formats.

\subsection{Applicability of LLSI}
We have shown that our approach using \ac{llsi} can detect a wide range of changes in the \ac{fpga} logic fabric configuration.
In the following, we discuss the practical applicability of \ac{llsi}.

\subsubsection{Chip Access}
For our approach, we need access to the silicon backside of the chip.
Since all \acp{fpga} used in this work are only available in flip-chip packages, this requirement can be easily met.
Moreover, due to performance, size, cost, and environmental compatibility reasons, chips are predominantly delivered in flip-chip packages~\cite{tong_advanced_2013}.
While many of such packages have a lid installed -- which we could easily remove for the PolarFire SoC -- there are also bare-die packages available, like the one of our Kintex-7 and UltraScale \acp{dut}.
Consequently, if a customer would like to have the opportunity to test the chip for \acp{ht} using an optical probing approach, he or she should choose a bare-die package to facilitate testing.
Thinning or polishing the silicon backside is not necessary for optical probing, as shown in this work.

\subsubsection{PCB Modifications}
In order to reach modulation frequencies of \SI{80}{\kilo\hertz} and higher, we had to replace the voltage regulator on the Kintex-7 and UltraScale \acp{dut} with an external one.
However, on the PolarFire \ac{dut}, we could leverage the on-\ac{pcb} regulator for the modulation, requiring no modifications on the \ac{pcb}.
Consequently, by using a suitable voltage regulator on the \ac{pcb}, there is no need to provide the modulated voltage from an external source.

During our investigations, we observed that a higher modulation of the supply voltage produces a clearer \ac{llsi} image, and consequently, a shorter pixel dwell time is sufficient.
Moreover, a higher modulation frequency can further reduce the pixel dwell time, leading to faster scan times.
The \ac{pcb} and the die interposer \ac{pcb}, however, are designed to compensate spikes and smooth undesired peaks and fluctuations of the supply voltage.
For this purpose, decoupling capacitors of different sizes are connected between the supply voltage rail and ground, effectively acting as low-pass filters.

To achieve the desired modulation amplitude of the power rail at frequencies above \SI{80}{\kilo\hertz}, we had to remove the decoupling capacitors of \SI{0.1}{\micro\farad} and larger from the \ac{pcb}.
Due to the existence of other capacitive and inductive elements in the circuit, a higher modulation frequency results in a lower modulation amplitude and, therefore, a lower \ac{llsi} signal level.
Consequently, there is a tradeoff between the noise ratio in the \ac{llsi} images, the scan time, and the electrical preparation of the \ac{dut}.
Due to practical reasons, we did not remove smaller capacitors.
Furthermore, we did not remove capacitors from the interposer \ac{pcb}, as there is no documentation on potential effects available.
Nevertheless, a device that is ready for use in a practical application must have installed all capacitors due to reliability and stability constraints.
One way to still enable the measurements required by our approach is the installation of jumpers or other switches on the \ac{pcb} to disable the capacitors on demand.

\subsubsection{Optical Stability}
In our experiments, we observed that the optical focus was slightly drifting during the \ac{llsi} measurements due to mechanical instabilities in the setup.
Since the \ac{llsi} signal heavily depends on the focus position, there are small differences between \ac{llsi} images that are not caused by design modifications.
However, the stability of our setup was sufficient to produce reliable and significant results for detecting malicious changes in the design.
Nevertheless, the image quality will improve if the mechanical stability is enhanced, for instance, by operating the setup in a tempered room and a shock-absorbing building.

\subsubsection{Optical Resolution}

The optical resolution of laser-assisted side-channel techniques has been discussed extensively by the research community in numerous publications, e.g., in~\cite{boit_ic_2016, lohrke_no_2016, tajik2017power, rahman_physical_2018, rahman_key_2020, stern2020sparta}.
We discuss the most important and new insights in the following.

Both \acp{fpga} used in this work were manufactured in \SI{28}{\nano\meter} and even \SI{20}{\nano\meter} technologies.
Although the minimum width of our setup's optical beam is around \SI{1}{\micro\meter}, it should be kept in mind that the technology size does distinguish neither the minimum size of a transistor nor the typical distance between transistors.
An important fact is that the laser scanner has a step size in the range of a few nanometers.
Therefore, while scanning with the laser over the \ac{dut}, the beam covers one specific point on the chip multiple times.
Consequently, if the beam covers multiple nodes of interest, the \ac{llsi} image shows a different position-dependent superposition of the same nodes at different adjacent pixel locations.
However, due to the Gaussian intensity distribution of the beam, it might still be possible to extract the logic state.
This explains why optical probing delivers meaningful results also on structures that are smaller than the beam diameter.

Moreover, a so-called \ac{sil} can be used to increase the optical resolution down to 250~\si{\nano\meter}~\cite{hamamatsu_nanolensshr_2015}, which is sufficient to resolve individual transistors in a 14\,\si{\nano\meter} technology~\cite{vonhaartm_optical_2015}.
Accordingly, Intel has shown that \ac{llsi} can be applied on very small devices, such as single inverters, on a test chip manufactured in a 14\,\si{\nano\meter} technology~\cite{niu_laser_2014}.

Even if it might not be possible to resolve single SRAM cells used for configuration storage in future technologies, the \acp{ff}, \acp{mux}, and  other pass transistors are influenced by the configuration and contribute to the \ac{llsi} image as well.
This is supported by the observation that even on the \SI{20}{\nano\m} FPGA, the different \ac{lut} configurations could be detected.
Furthermore, typical \acp{ht} in benchmarks alter the design by inserting or modifying multiple logic gates or \acp{ff}, resulting in huge changes, which we could detect reliably.

\section{Conclusion \label{sec:conclusion}}
Dormant hardware Trojans that introduce only tiny malicious hardware modifications pose a severe threat in security-critical applications.
In this work, we have demonstrated a detection approach for dormant \acp{ht} using the laser-assisted optical probing method \ac{llsi}.
By modulating the power supply of the chip, even inactive logic is visible on the logic snapshots.
By awakening the potential Trojan in this way, no malicious modification of the \ac{fpga}'s configuration stays undetected.
We have demonstrated that our approach is applicable to recent SRAM- and flash-based \acp{fpga} on the market in a non-invasive manner.
It did not make a significant difference whether the \acp{fpga} were manufactured in a \SI{28}{\nano\m} or \SI{20}{\nano\m}  technology.
Finally, we have explained why our framework should also be suitable for detecting stealthy \acp{ht} on \acp{asic}.

\ifthenelse{\boolean{cameraready}}{	
\section*{Statements and Declarations}
\textbf{Funding}\enskip\enskip
The authors from Technische Universität Berlin have been supported in part by the Einstein Foundation (EP-2018-480), and in part by the Deutsche Forschungsgemeinschaft (DFG -- German Research Foundation) under the priority programme SPP 2253, grant number 439918011.
For the author of Worcester Polytechnic Intitute, the effort was sponsored in part by NSF under grant number 2117349.

\smallskip
\noindent
\textbf{Data Availability}\enskip\enskip
The datasets generated during and analyzed during the current study are available from the corresponding author on reasonable request.

\smallskip
\noindent
\textbf{Acknowledgement}\enskip\enskip
The authors would like to acknowledge Hamamatsu Photonics K.K. Japan and Germany for their help and support on the \mbox{PHEMOS} system.
}{}

\FloatBarrier
\bibliographystyle{./ACM-Reference-Format}
\bibliography{./bib/IEEEabrv,./bib/ACMabrv.bib,./bib/references}
\FloatBarrier

\end{document}